\documentclass[preprint,12pt]{elsarticle}

\usepackage{amssymb}          
\usepackage[latin1]{inputenc}                                
\usepackage{color}                                           
\usepackage{array}                                           
\usepackage{longtable,ltcaption}                                       
\usepackage{calc}                                            
\usepackage{multirow}                                        
\usepackage{hhline}                                          
\usepackage{ifthen}                                          

\journal{IJROBP}

\begin{document}

\begin{frontmatter}

\title{3D conformal planning using low segment multi-criteria IMRT optimization}

\author{Fazal Khan and David Craft}
\address{Massachusetts General Hospital \\ Harvard Medical School \\ Boston, MA}

\begin{abstract}
\noindent {\bf Purpose:} To evaluate
automated multicriteria optimization (MCO) -- designed for intensity modulated radiation
therapy (IMRT), but invoked with limited segmentation -- to efficiently produce
high quality 3D conformal treatment (3D-CRT) plans. 

\noindent {\bf Methods:} Ten patients
previously planned with 3D-CRT were replanned with a low-segment
inverse multicriteria optimized technique. The MCO-3D plans used the
same number of beams, beam geometry and machine parameters of the corresponding
3D plans, but were limited to an energy of 6 MV. The MCO-3D plans
were optimized using a fluence-based MCO IMRT algorithm and then, after
MCO navigation, segmented with a low number of segments. 
The 3D and MCO-3D
plans were compared by evaluating mean doses to individual organs
at risk (OARs), mean doses to combined OARs, homogeneity indexes (HI),
monitor units (MUs), physician preference, and qualitative assessments of planning time and
plan customizability. 

\noindent {\bf Results:} The MCO-3D plans significantly reduced the OAR mean doses and 
monitor units while maintaining good coverage and homogeneity of target volumes. MCO 
allows for more streamlined plan customization. All MCO-3D plans were preferred by 
physicians over their corresponding 3D plans.

\noindent {\bf Conclusion:} High quality 3D plans can be produced using
IMRT optimization technology, resulting in automated field-in-field type plans with
good monitor unit efficiency. Adopting this technology in a clinic could streamline 
treatment plan production.

\end{abstract}

\end{frontmatter}

\section{Introduction}

3D planning, also known as forward planning, is a standard approach for
delivering conformal radiotherapy to a variety of cancers. The simplicity,
low cost, low maintenance and well documented outcomes of 3D planning
have made it the preferred choice for many disease sites. 
In 3D planning, dose distribution changes are made as
a direct result of the planner manually modifying various treatment
parameters such as field shapes, beam weights, beam modifiers, 
and dose normalization.

The 2000s saw a growing interest in intensity modulated
radiation therapy (IMRT), a computer optimized method of delivering
radiation \cite{Webb2003} which modulates radiation from each
field through the use of a multi-leaf collimator (MLC), permitting greater
conformality and better OAR sparing. However 
IMRT comes with its own challenges including greater susceptibility to 
motion \cite{imrtmotion,imrtmotion2}, more complex
dosimetry, potentially higher monitor units and treatment time \cite{imrtmu}, increased 
quality assurance (QA) complexity
and greater machine wear-and-tear \cite{imrtweartear}.
IMRT can cost anywhere from 1.5 to four times the
amount of a 3D plan \cite{cost1,cost2,cost3}, 
and resistance from insurance companies to reimburse
for IMRT adds to the persistence of 3D conformal therapy in clinics
worldwide (e.g. \cite{reimburse}). Some disease sites, notably prostate and head-and-neck,
have moved to IMRT planning for the majority of their cases, but many
common sites such as breast and lung remain in the 3D planning realm.

Due to the manual manipulation required in 3D planning it can be
time-consuming to find a desirable dose distribution.
Also, once a plan is created
there is no way of confirming whether the plan is fully optimized.
By `fully optimized' we refer to a plan where any improvement of one criteria --
eg. homogeneity or OAR dose -- must come
at the expense of worsening another planning criteria. This requirement
is known as Pareto optimality. The set of Pareto optimal plans is
called the Pareto surface and exploring this surface 
has become a valuable technique for
IMRT planning \cite{Craft2011,fredriksson2013}. MCO for IMRT allows
the planner to smoothly navigate through all the generated plans by
mixing individual plan fluences, allowing for a quick exploration of
possible plans. Comparison studies of MCO
and standard planning for IMRT have shown MCO can significantly minimize
the time needed to generate a plan,  
while producing plans preferred by the physicians \cite{Craft2011}.
 
Since Pareto navigation hinges on the ability to average multiple
plans, IMRT (as well as intensity modulated proton therapy) is an
ideal modality since fluence maps can be averaged, which leads to
the averaging of dose distributions. Because common 3D conformal sites
only require a small amount of intensity modulation, 
we hypothesized that using fluence map based MCO and
a low number of segments would allow us to use the MCO-IMRT
planning technique to generate high quality 3D conformal plans.
In this way, 
we get the best of the two worlds: the relative simplicity of
3D plans with the power of numerical optimization that comes with IMRT.
Empowered by an MCO Pareto surface, the planner and physician can rapidly explore
dose tradeoffs. Lastly, because we are using low segments, no physical wedges, and
an efficient optimizer, we reduce MUs, retain plan robustness, reduce patient treatment
time and reduce the need for patient-specific QA.

\section{Methods and materials}

\subsection{Case selection and structure definitions}

Ten recently planned patients of various disease sites
(breast, brain, lung, abdomen and pelvis) were selected from our institutional
clinical database. These sites were chosen
to show the technique across a wide range of 3D planned areas
of the body. 

The original physician-drawn target volumes and OARs as well as the
original dosimetrist-generated target expansion volumes were used
to plan and evaluate all plans.
In the MCO-3D cases additional planning structures were created to
help guide the optimizer. Each MCO-3D plan contained a structure expanded
from the PTV radially to the edge of the CT scan but only 4cm superiorly
and inferiorly, effectively creating a wide cylindrical volume. This
structure was termed `falloff' and was used to promote dose conformality.
Some MCO-3D plans used an additional 2-3cm wall expansion around
the PTV, called `PTVwall', for additionally sharpening high-dose
conformality.

For the breast cases, the physician-drawn targets include only the
breast contour and the seroma. For planning purposes, an artificial
PTV was created by taking the breast contour and contracting it from
the edge of the CT scan by 2mm. This was to prevent the optimizer
from aggressively trying to deliver full dose near the patient's skin.

For all patients we also created a structure called `total OAR' which
was the union of all the OARs, in order to evaluate the overall mean
dose to OARs.

\subsection{Planning Parameters}

XiO (v4.4; Elekta, Stockholm, Sweden) was used to plan and 
calculate all original 3D plans. We selected
recent plans from the clinical database, thus the planning strategy
for the 3D conformal plans was our standard clinical procedure.
RayStation (v2.5; RaySearch Laboratories, Stockholm, Sweden) 
was used to optimize and calculate all MCO-3D plans. The MCO-3D
plans matched the original 3D plans in terms of machine
(Varian or Elekta), number of beams and beam geometry. The
MCO-3D plans did not use any beam modifiers (i.e. wedges),
and were limited to an energy of 6 MV. All MCO-3D plans followed
the original 3D plan prescription dose and fractionation schemes.

Pareto surface-based MCO uses the classical optimization paradigm
of objectives and constraints. Constraints are criteria which cannot
be violated while objectives are functions
which are minimized or maximized subject to the constraints. All OARS
were assigned a `minimize the equivalent uniform dose (EUD)' objective \cite{geud,wuEUD}.
EUD for an organ with $n$ equi-sized voxels, each receiving $d_{i}$
dose, is given by 
\begin{equation}
\mathrm{EUD}=\left(\frac{1}{n}\sum_{i}d_{i}^{a}\right)^{1/a},
\end{equation}
where $a$ is a parameter generally chosen to be greater than or equal
to 1. If $a=1$, the EUD is the mean dose to the organ. As $a$ is
increased, the function is weighted more heavily towards larger doses.
In the limit of $a\rightarrow\infty$, the EUD approaches the maximum
dose of the organ. The EUD is a convex function which makes it appealing
for optimization purposes. We use $a=2$ which is a standard approach
to controlling both the mean dose and the hotspots.

For the falloff structure we use the `dose-falloff' objective. This objective
penalizes doses outside the target by specifying a desired dose falloff
rate (as a function of distance to the target). Voxels which violate
this dose falloff are penalized quadratically based on their deviation.

In the cases where the PTVwall structure was used, this structure
was given an EUD objective with an $a$ value ranging from 20-30 to
penalize high doses. The target volume (PTV in all cases) was given
both objectives and constraints. The target objectives consisted of
a minimum dose objective, which is a quadratic penalty on voxel underdosage,
and a uniform dose objective (the standard two-sided quadratic penalty),
both at the prescription dose. The target constraints included a dose-volume
constraint of at least 95\% of the volume receiving the prescription
dose as well as a minimum target dose of 95\% of the prescription.
Lastly, a constraint was given on the falloff volume as a max dose
equal to 105\% of the prescription dose. This served to limit the
global maximum dose of the plan.

Once the objectives and constraints are entered, RayStation
computes a set of Pareto optimal plans. This begins with anchor plans,
which optimize each objective individually, while respecting the 
constraints. Once all anchor plans are generated, RayStation
creates plans where two or more objectives are simultaneously
optimized. These auxiliary plans help to enrich the Pareto surface
for better navigation. The total
number of plans computed is a user-defined parameter. In this study
we used 4$N$ plans for each Pareto surface, where $N$ is the number
of objectives defined. After navigation, RayStation's direct machine parameter 
optimization is invoked which creates a deliverable plan. In this step the system
determines MLC segment shapes and weights to best create the navigated-to dose.

\subsection{Determination of number of segments}

Since the IMRT module of RayStation does not support higher energies or wedges,
we opted to allow the MCO-3D plans a few additional segments to allow the MCO-3D plans to
compete more fairly with the standard 3D conformal plans, which utilize wedges, 
higher energies, and field-in-fields (FIFs). RayStation
allows the planner to constrain the maximum number of segments used
for the segmentation of a fluence optimized plan, and it automatically
determines which beams will have additional segments, with the stipulation
that each beam gets at least one segment. 
We determined the number of segments using the following:
\begin{enumerate}
\item One segment per unique beam angle in the original 3D plan 
\item One segment per field-in-field used in the original 3D plan.
\item We add the number of fields using higher energies (HE) than 6 MV to the
number of wedges (W) used to get ``HE+W'', then add additional segments to
the MCO-3D plan based on the following.

\begin{itemize}
\item If the HE+W = 1-2, we add 1 additional segment 
\item If the HE+W = 3-5, we add 2 additional segments 
\item If the HE+W = 6+, we add 3 additional segments 
\end{itemize}
\end{enumerate}

\subsection{Pareto surface navigation and final plan selection}

Each MCO-3D plan was navigated to reduce OAR doses while meeting one
of the following criteria: 
\begin{itemize}
\item the plan met or exceeded the original 3D plan's coverage at prescription
dose 
\item 95\% of the PTV volume received prescription dose (our clinical standard). 
\end{itemize}
After navigation and segmentation, normalization (scaling) was used
to achieve coverage, if necessary.

\subsection{Evaluation}

Once all MCO-3D plans were completed they were evaluated for clinical
acceptability. We evaluated each plan by comparing individual OAR
mean dose, total OAR mean dose, MU and homogeneity indexes. The homogeneity
index (HI) is defined as:
$$
\mathrm{HI} =  \frac{D_5-D_{95}}{D_p} 
$$
\noindent where $D_5$ is the dose
to 5\% of the PTV, $D_{95}$ is the dose to 95\% of the PTV, and $D_p$ is the
prescription dose. A perfectly homogeneous PTV dose would
have $D_5$, which measures hotspots, equal to $D_{95}$, which measures the
cold spots. Therefore the best achievable value for HI is 0.

All dose computations
were done with the in-use clinically commissioned systems.
The original 3D plans and the MCO-3D plans were exported to
MimVista (Version 6.0, Cleveland, OH) for evaluation, in order to
eliminate inherent differences
in DVH computations by the two planning systems. 

\section{Results}

\subsection{Case descriptions}

We describe each case in terms of the site, prescription,
beam energies and geometries, and techniques used in the original
3D plan. For each case we state any significant difference between
the original 3D plan and the MCO-3D plan regarding OAR sparing and homogeneity. We also indicate any
significant difficulties encountered in the MCO-3D planning including
hotspots occurring outside of the PTV, dose streaking, and maintaining
an acceptable homogeneity index within the PTV. Dose and MU comparison
data for all cases are summarized in Table~1. We selected four of the cases
to display the dose distribution and DVH comparisons. We selected these to show 
a range of results (we did not select the cases which yielded the `best' results for
the MCO-3D planning, we selected a representative set).

\medskip
\noindent{\bf Brain cases}

Case 1 was a posterior fossa tumor prescribed to 20 Gy.
A 4 field X-shaped beam arrangement was used. The original 3D plan
utilized all 10 MV beams as well as four wedges and two FIFs. The
only OARs drawn were the cochleas. The MCO-3D spared both cochleas
significantly more than the original 3D plan and created
a much steeper dose gradient outside the PTV. This OAR sparing
came at the price of a reduction in homogeneity, with an HI of .07
compared to .03 for the original plan. The PTVwall structure
was helpful in controlling hotspots outside the PTV volume.

\begin{figure}[h!t]
\centering
\includegraphics[trim=0 100 0 50,clip,width=15cm]{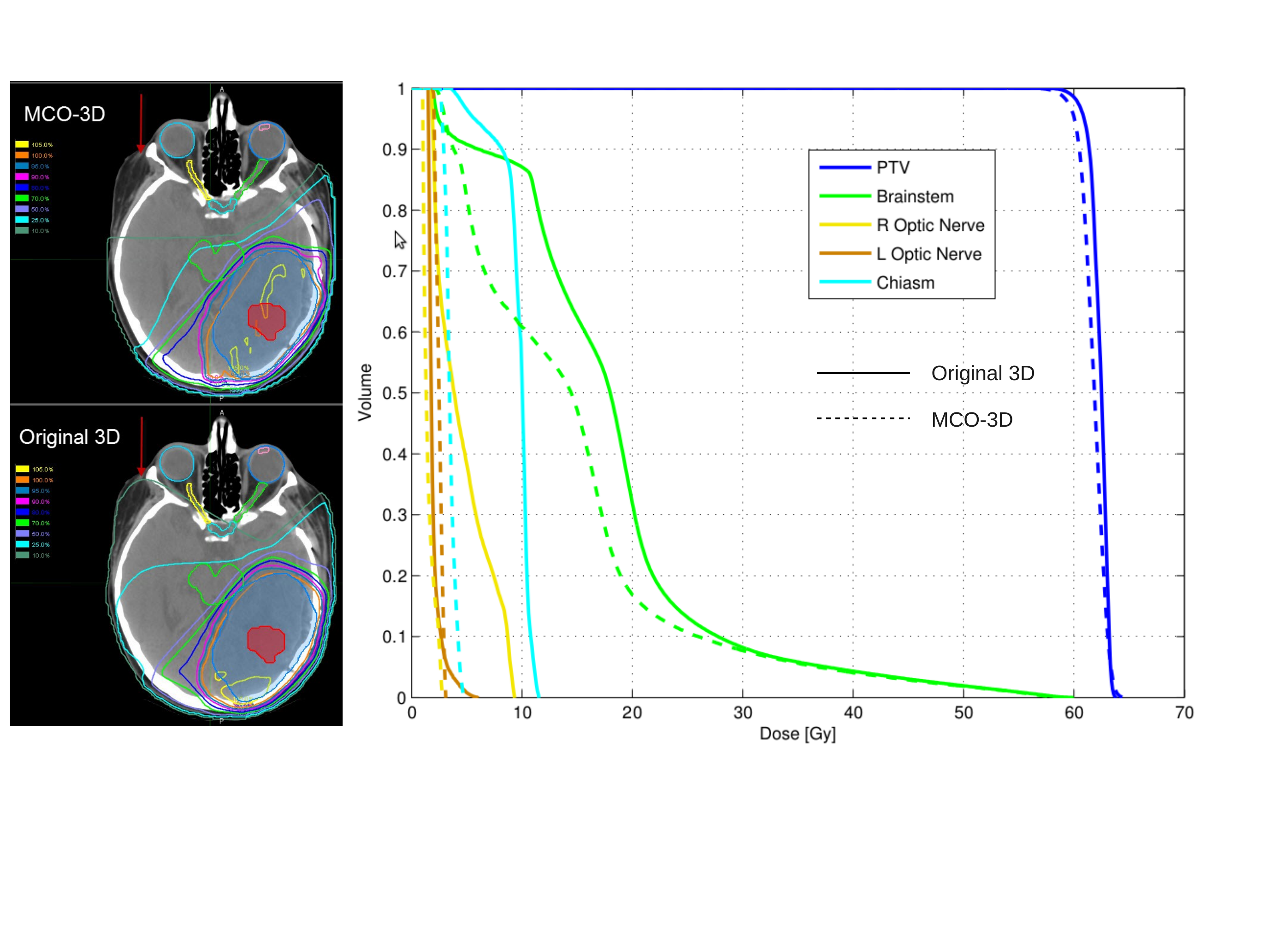}
\caption{Axial dose distribution and DVH comparison for brain case 2. The red arrows 
highlight MCO-3D pushing low dose away from critical organs.}
\label{brain2}
\end{figure}

Case 2 was a left parietal tumor prescribed to 60 Gy. A 5-field beam
arrangement was used: four coplanar beams and one superior vertex
beam. The original 3D plan used 6 MV for all coplanar beams and 10
MV for the superior vertex beam. Three wedges were also used. The
most notable sparing observed in the MCO-3D plan were in the chiasm,
R. cochlea and R. optic nerve; each of their mean doses were lowered
by a factor of three. There were very small increases in mean dose
for the L. optic nerve and the L. cochlea. As in brain case 1, this
significant OAR sparing came at the price of a small increase in the
homogeneity index, from .04 to .06. The comparison between the original
3D plan and the MCO-3D plan for brain case 2 is shown in Figure \ref{brain2}.

\medskip
\noindent{\bf Breast cases}

Case 1 was a right sided breast to 50 Gy. This patient had
a small breast and the original 3D plan employed two open tangent 6 MV
fields. The HE+W number was zero, thus the MCO-3D plan utilized only
two segments total (one segment for each field). The MCO-3D plan was
able to drastically improve breast coverage at prescription
dose (15\% increase) while simultaneously reducing lung dose. Hotspots and the global
maximum were kept nearly identical to the original 3D plan. The homogeneity
index was improved from .31 to
.18, due to the increase in coverage without an increase in hotspots.

\begin{figure}[h!t]
\centering
\includegraphics[trim=0 200 0 30,clip,width=15cm]{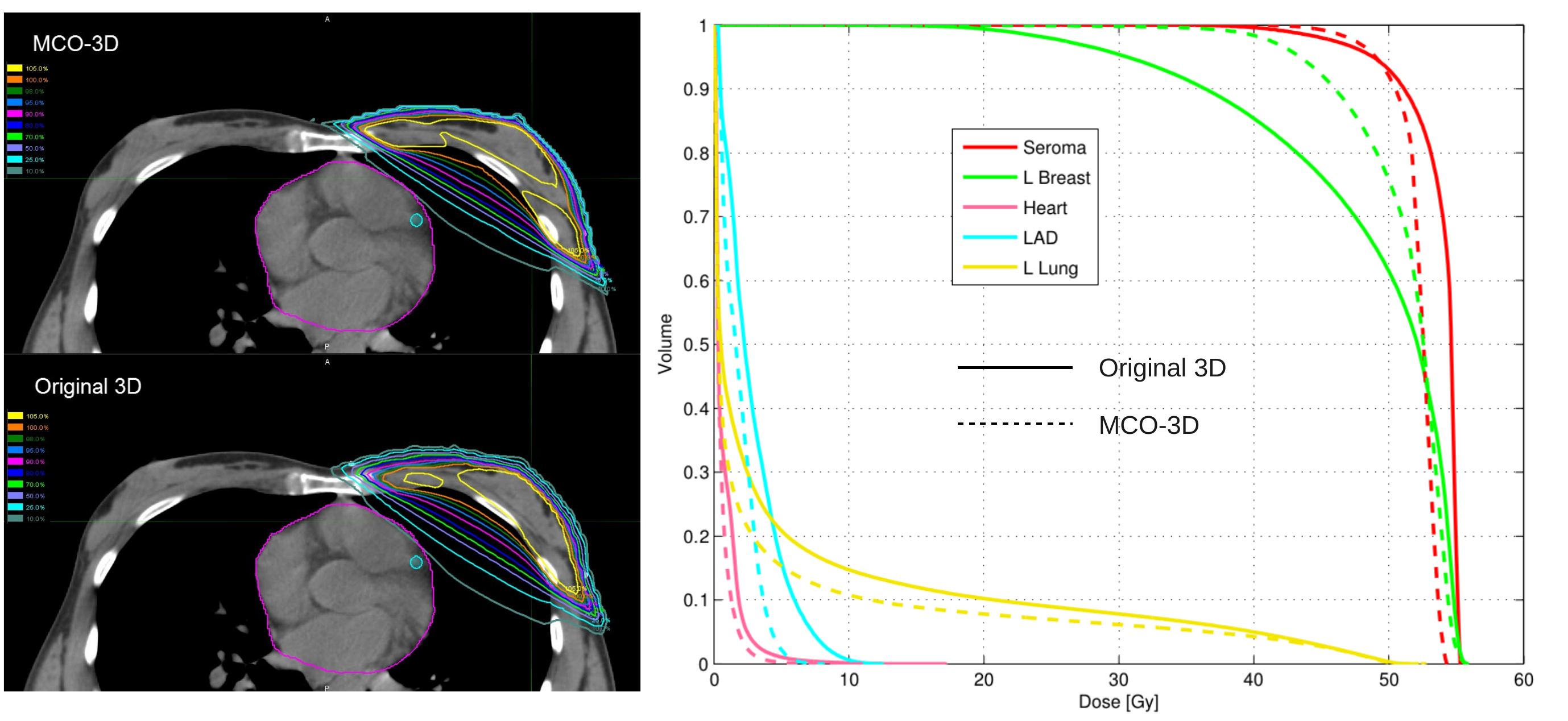}
\caption{Axial dose distribution and DVH comparison for breast case 2.}
\label{breast2}
\end{figure}

Case 2 was a left sided breast, also to 50 Gy, which allowed us to
test our technique with a case involving the whole heart, left ventricle
and left anterior descending (LAD) artery in addition to the lung.
The original 3D plan utilized two open tangents and two field-in-fields.
10 MV was used on the medial side while 6 MV was used on the lateral
side. The plan's HE+W number was two, therefore we used one additional
segment beyond the original number of parent fields and FIFs, for
a total of five segments. Similar to case 1, the MCO-3D demonstrated 
superior breast coverage while simultaneously
lowering whole heart, left ventricle, LAD and lung dose. Once again,
hotspots remained similar to the original 3D plan and the homogeneity
index improved from .43 to .21. The plan comparison 
is shown in Figure \ref{breast2}.

\medskip
\noindent{\bf Thoracic cases}

\begin{figure}[h!t]
\centering
\includegraphics[trim=0 120 0 30,clip,width=15cm]{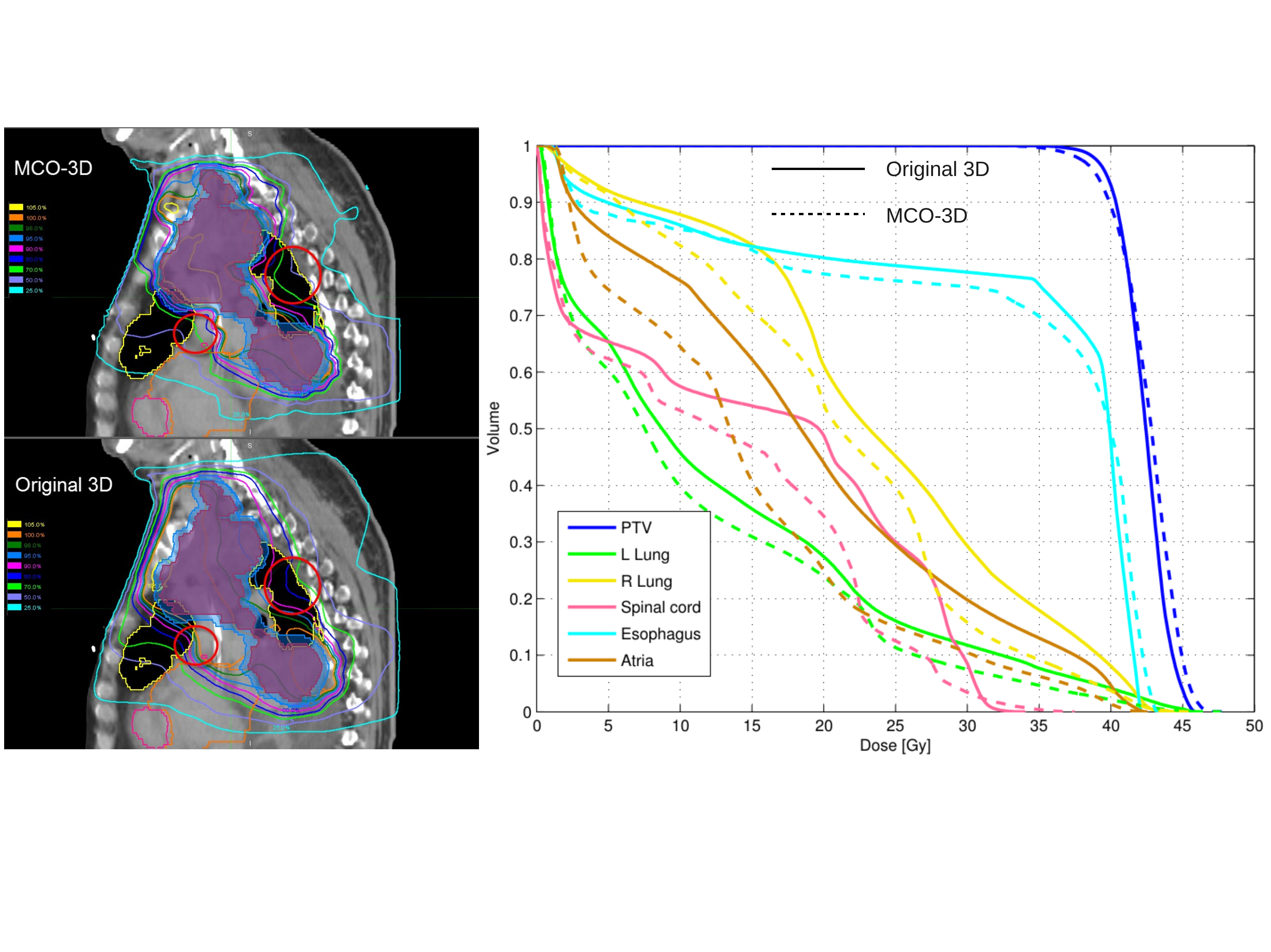}
\caption{Axial dose distribution and DVH comparison for thoracic case 1. The circled regions indicate
areas where MCO-3D clearly sculpts the dose distribution to conform to the PTV.}
\label{thor1}
\end{figure}

Case 1, the most challenging of the ten cases, 
was a large lung volume with a prescription dose
of 42 Gy. The GTV was very extensive and branched into many nodal chains
throughout the thorax. The PTV volume was 957 cubic centimeters. 
The original 3D plan used five coplanar fields -- one
anterior and four obliques. 10 MV was used for all beams as well as
four wedges. The MCO-3D reduced every OAR at
the expense of a small decrease in homogeneity. 
The plan comparison is shown in Figure~\ref{thor1}.

Case 2 was an esophagus prescribed to 55 Gy. The original
plan employed a four field posterio-lateral beam arrangement. Wedges
and 10 MV beams were used on all fields, leading to seven segments
total for the MCO-3D plan. The MCO-3D lowered the mean dose
of every OAR although not as significantly as in thoracic Case 1.
There was no significant change in homogeneity however MCO-3D did
encounter difficulties with high dose building up near the skin. A
2cm inner wall contour created from the external contour was necessary
to help control this.

\medskip
\noindent{\bf Abdomen cases}

Case 1 was a pancreas volume prescribed to a dose of 30
Gy. A 4-field conformal beam arrangement was used. All beams were
15 MV and three wedges were used in the original plan. The MCO-3D plan
lowered the mean dose of most OARs. The small bowel and R. kidney were lowered more
significantly while the spinal cord dose was increased. There was
a small homogeneity index increase for the MCO-3D plan, from
.06 to .07.

Case 2 was a larger pancreas volume prescribed to 45 Gy. The original 3D plan used five beam
angles, with 15 MV for all
beams, and two wedges. The MCO-3D significantly lowered all
OAR mean doses. The homogeneity index rose from .05 to .08. Initially,
some dose streaking was encountered in the MCO-3D plan. This case successfully 
showed MCO-3D's ability to treat larger and deeper seated abdominal volumes.

\medskip
\noindent{\bf Pelvis cases}

\begin{figure}[h!t]
\centering
\includegraphics[trim=0 200 0 30,clip,width=15cm]{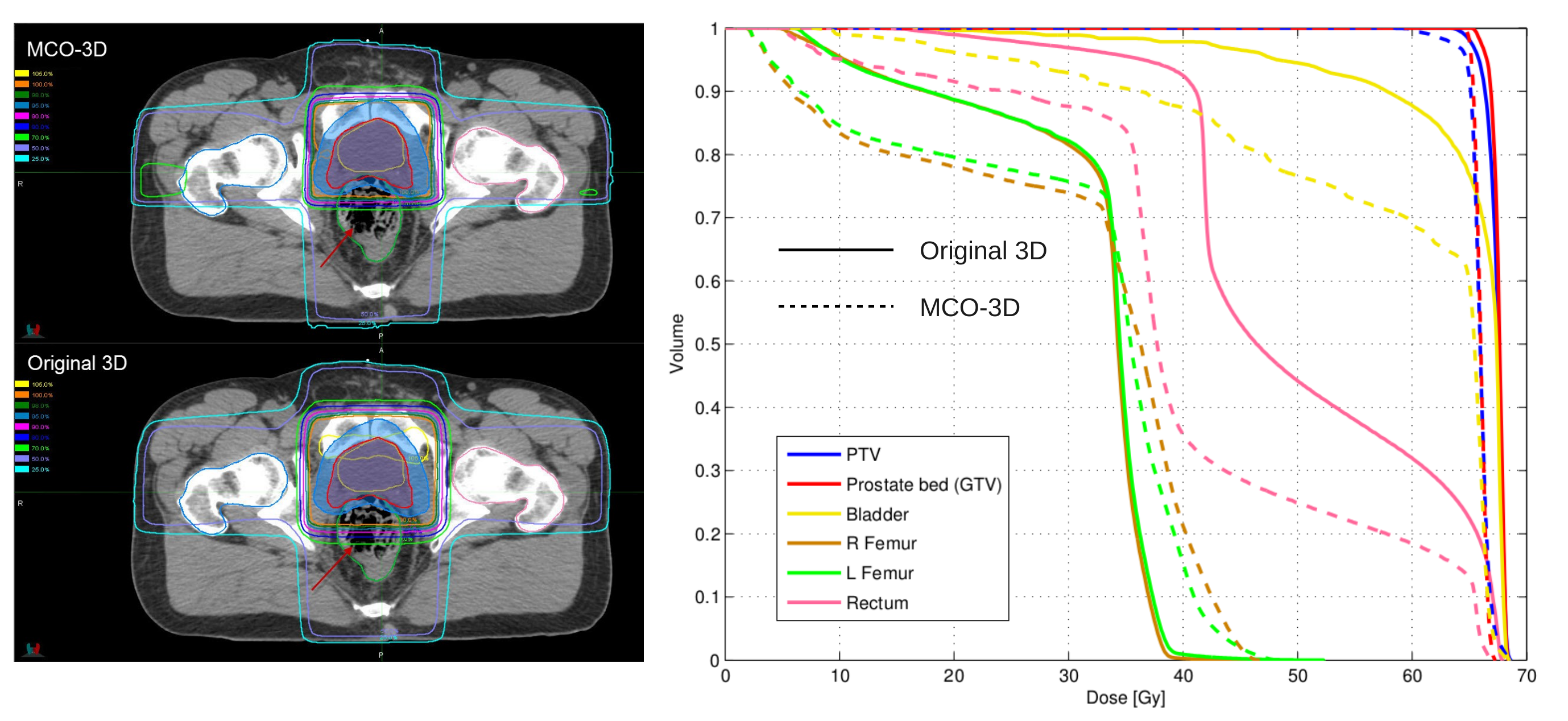}
\caption{Axial dose distribution and DVH comparison for pelvis case 1. The red arrows show that
the MCO-3D plans are able to push dose out of the rectum.} 
\label{pelv1}
\end{figure}

Case 1 was a standard four-field box prostate fossa prescribed
to 64.8 Gy. The original plan used 10 MV on all beams except the
AP beam which used 6 MV. The bladder and rectum were spared
very well in the MCO-3D plan. Femoral head mean doses were also lower
in the MCO-3D plan. Higher lateral entrances doses occurred with the 
MCO-3D plan. Homogeneity indexes remained the same as the original
3D plan. The comparison between the original
3D plan and the MCO-3D plan for pelvis case 1 is shown in Figure \ref{pelv1}.

Case 2 was a three-field bladder prescribed to 39.6 Gy. The original
3D plan used two oblique laterals and one right anterior oblique beam.
10 MV was used for all three fields and two wedges were also used.
The MCO-3D plan was able to reduce the mean dose to all OARs, especially 
the femoral heads. No significant difficulties were encountered
with the MCO-3D plan, however the homogeneity index increase in this
plan was the greatest, from .05 to .09.

\begin{table}[p]
\centering 
\includegraphics[trim=55 20 40 25,clip,width=15.5cm]{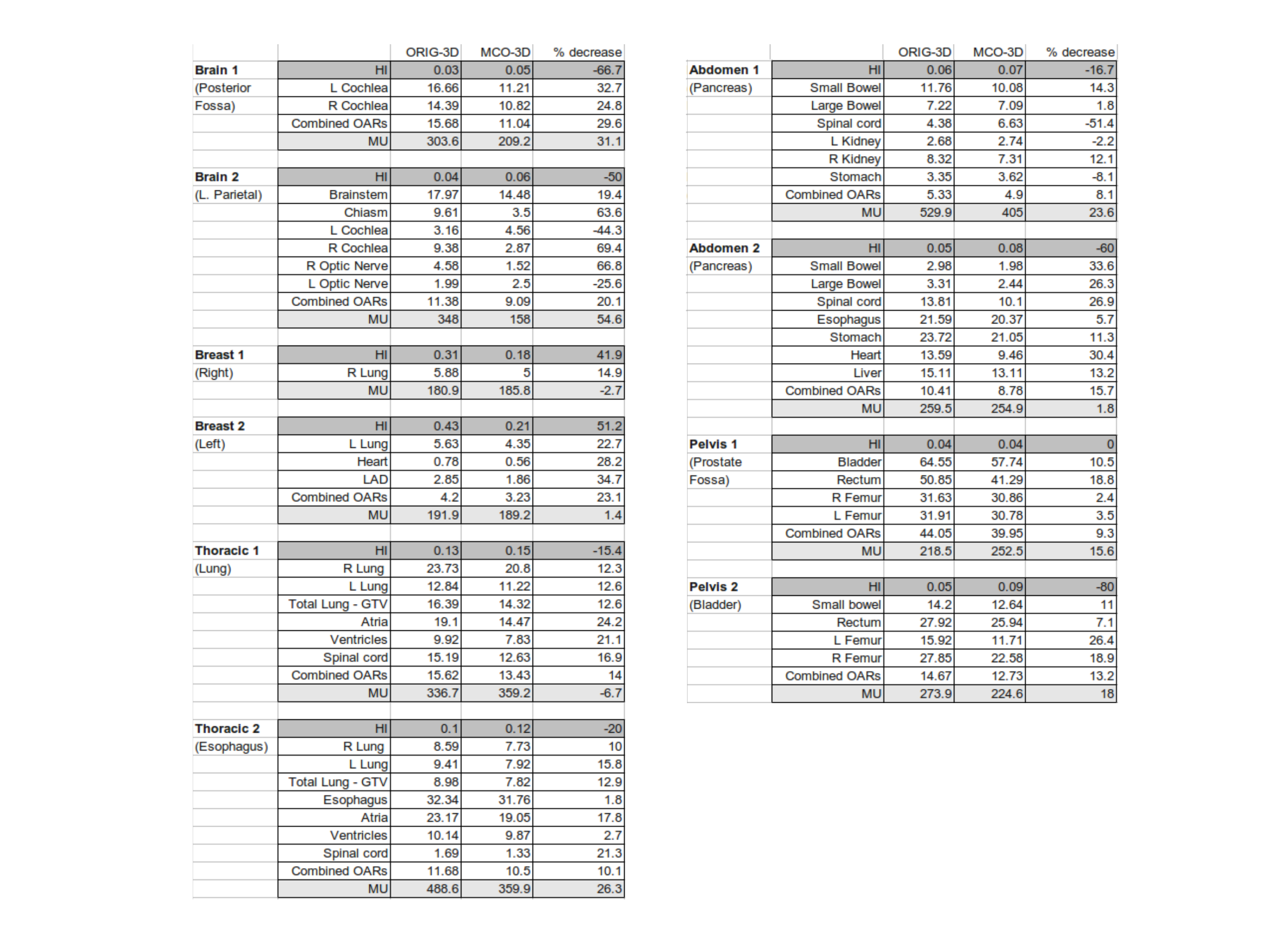}
\caption{Dosimetric and monitor unit (MU) comparisons for all ten patient cases. 
For OARS, mean dose in Gy is reported, and for the target, homogeneity index HI, 
as defined in the text, is reported.}
\end{table}


\subsection{Summary of plan comparisons: doses, MUs, physician preference, and planning}

In every MCO-3D plan, the majority of OAR mean doses were lowered compared to
the original 3D plan. In some cases the MCO-3D plan gave one or two
OARs a greater mean dose, however this slight increase was overcome
by lowering others OARs even more. In the cases where some OARs are lower while 
others are higher it can sometimes be difficult
to evaluate whether there was an overall improvement in mean OAR dose.
In order to quantify the overall reduction of mean dose to organs,
we evaluated the mean dose to the total OAR structure.
In each case the MCO-3D had a lower total OAR mean dose, see Table~1. 

After all the MCO-3D plans were generated they were coupled with their original 
3D plan and sent back to the treating
physician to ask which plan they preferred. {\em In all ten cases the physicians selected
the MCO-3D plans over the original 3D plans.}

On average, fewer MUs were required by the MCO-3D plans. The average
MU of the original 3D plans was 313 while the average MU of the
MCO-3D plans was 259, a 17\% decrease. This finding defeats the
long held belief that inverse planning necessarily produces greater
MUs than forward 3D planning: it depends on the number of segments being used. 
The brain cases had the most significant total reduction in MU, with a combined total of 
284 fewer MUs for MCO-3D. The least change in MU were in the breast cases.

Treatment planning times for our MCO-3D plans were
similar to traditional 3D planning (although the bulk of the MCO-3D
planning time was used in computing Pareto surfaces, a process which
has been improved in later versions of RayStation, with more increases expected by 
moving to a distributed computing environment \cite{bokrantz2}). 
The use of higher energy beams in MCO-3D will also help the planning time by 
alleviating dose streaking and improving homogeneity.

\section{Discussion and Conclusions}

Photon treatments are typically classified as either 3D conformal
plans or IMRT plans. It is more useful for understanding treatment
plan optimization to think of these modalities as lying along a
continuous span of treatment possibilities, from
simple to complex \cite{meng2010unified}. Although somewhat counterintuitive, IMRT treatment
planning is sometimes easier than 3D conformal since
numerical optimization can be used to find optimal solutions, and planning
software and computation power have improved dramatically since the early days of IMRT.
Considering this, we speculated that if IMRT optimization was used -- in particular MCO -- 
we might be able to derive good 3D solutions from the selected IMRT
plan if that plan would naturally not require too much intensity modulation.
For example, IMRT optimization applied to a spherical tumor might
yield relatively flat beam profiles which could then be delivered
with open 3D conformal fields. In the ten cases we examined, 
using the IMRT optimizer to produce 3D plans led to
significant OAR sparing at the cost of a small decrease in target
homogeneity, and the treating physicians unanimously preferred the MCO-3D
plans to the original 3D plans.

A single treatment planning system used for both 3D conformal planning
and IMRT planning would be beneficial from a training and quality assurance
perspective. Fewer systems means an overall operation that
is easier to monitor and less prone to error \cite{systemsafety,leveson}.

While this is not the place for a full discussion of the insurance
and billing differences between 3D and IMRT, which is related to
the historical difficulty of planning and delivering IMRT and 
the fact that IMRT plans are typically quality assured (QAed)
by measuring the plan dose on a phantom while 3D plans are not, we
suggest that clinics who adopt the MCO-3D method presented herein discuss considering
such plans as 3D. Our view on this issue is that since there is a
spectrum of plan complexity between 3D conformal and IMRT, this spectrum
should be realized more fully in the clinic: a plan should
be as complex as necessary to achieve a desired level of dose quality.
QA procedures should be standardized, and should take the form of
independent software -- such as a Monte Carlo system -- in order to
automatically verify all plans \cite{monteVerify1,monteVerify2},
thus eliminating the QA distinction between 3D and IMRT plans.

In modern clinics, IMRT has become the clinical standard for sites
which most strongly benefit from being able to shape the dose distribution
to avoid nearby OARs. However, all sites could benefit from some intensity
modulation, which is why 3D conformal therapy has evolved to include
FIFs, wedges and higher
energies. These advanced technologies provide dose control similar
to IMRT \cite{IMRTBREAST2011}. For all of the sites studied in this
paper, IMRT has been explored \cite{reimburse,IMRTesoph, bladder, brain} 
and is often used, but 3D remains a common modality for treatment. 
One likely reason for this is that IMRT may often seem
overly complex, more costly and less efficient for the treatment goals
in mind. Our technique on the other hand depends on the idea that
a little intensity modulation goes a long way, as brought to light
by the many studies which point out the vastly diminishing returns
one gets from adding more complexity (larger MU and more segments)
to a plan \cite{craft-spg,SunXia04,WebbEtAl1998,AlberNuesslin2000}.
Our planning method allows the customizeability and dosimetric benefits
of MCO-IMRT with the simple and robust delivery of 3D conformal therapy.

\bigskip
\bigskip

\noindent {\bf Acknowledgment:} The authors thank Tarek Halabi, Thomas 
Bortfeld, and Stephen Zieminski for their valuable input
during the preparation of this manuscript. 

\noindent  
\bibliographystyle{elsarticle-num}
\bibliography{all}

\end{document}